\documentclass[12pt,preprint]{aastex}

\slugcomment{Accepted for Publication in The Astrophysical Journal}

\shorttitle{New Search for CO in HD\,209458b}
\shortauthors{Deming et al.}

\begin{document}

\title{A New Search for Carbon Monoxide Absorption in the Transmission Spectrum
of the Extrasolar Planet HD\,209458b\altaffilmark{1}}

\author{Drake Deming\altaffilmark{2}, Timothy M. Brown\altaffilmark{3,4}, 
David Charbonneau\altaffilmark{5}, Joseph Harrington\altaffilmark{6},\\ 
and L. Jeremy Richardson\altaffilmark{7}}

\altaffiltext{1}{Data presented herein were obtained at the W.M. Keck Observatory, which is 
operated as a scientific partnership among the California Institute of Technology, 
the University of California and the National Aeronautics and Space
Administration. The Observatory was made possible by the generous financial 
support of the W.M. Keck Foundation.}
\altaffiltext{2}{NASA's Goddard Space Flight Center, Planetary Systems Branch,
Code 693, Greenbelt MD 20771, email: ddeming@pop600.gsfc.nasa.gov}
\altaffiltext{3}{High Altitude Observatory/National Center for Atmospheric Research, 
3450 Mitchell Lane, Boulder CO 80307, email: timbrown@hao.ucar.edu}
\altaffiltext{4}{The National Center for Atmospheric Research is sponsored by the 
National Science Foundation.}
\altaffiltext{5}{California Institute of Technology, 105-24 (Astronomy), 
1200 E. California Blvd., Pasadena, CA 91125, Present address:
Harvard-Smithsonian Center for Astrophysics, 60 Garden St., MS-16,
Cambridge, MA 02138, email: dcharbonneau@cfa.harvard.edu}
\altaffiltext{6}{Center for Radiophysics and Space Research, Cornell University, 
 326 Space Sciences Bldg., Ithaca, NY 14853-6801, email: jh@oobleck.astro.cornell.edu}
\altaffiltext{7}{National Research Council, NASA's Goddard Space Flight Center, 
Infrared Astrophysics Branch, Code~685, Greenbelt MD 20771, 
email: Lee.J.Richardson.1@gsfc.nasa.gov}

\begin{abstract}

We have revisited the search for carbon monoxide absorption features
in transmission during the transit of the extrasolar planet
HD\,209458b.  In August-September 2002 we acquired a total of 1077
high resolution spectra ($\lambda / \delta \lambda \sim 25,000$) in
the K-band (2 $\mu$m) wavelength region using NIRSPEC on the Keck~II
telescope during three transits. These data are more numerous and of
better quality than the data analyzed in an initial search by
\citeauthor{brown02} Our analysis achieves a sensitivity sufficient to
test the degree of CO absorption in the first overtone bands during
transit, based on plausible models of the planetary atmosphere.  We
analyze our observations by comparison to theoretical tangent geometry
absorption spectra, computed by adding height-invariant {\it ad hoc}
temperature pertubations to the model atmosphere of
\citeauthor{sud03}, and by treating cloud height as an adjustable
parameter.  We do not detect CO absorption.  The strong 2-0 R-branch
lines between 4320 and 4330 cm$^{-1}$ have depths during transit less
than 1.6 parts in $10^{4}$ in units of the stellar continuum
(3$\sigma$ limit), at a spectral resolving power of 25,000.  Our
analysis indicates a weakening similar to the case of sodium,
suggesting that a general masking mechanism is at work in the
planetary atmosphere.  Under the interpretation that this masking is
provided by high clouds, our analysis defines the maximum cloud top
pressure (i.e., minimum height) as a function of the model atmospheric
temperature.  For the relatively hot model used by
\citeauthor{charb02} to interpret their sodium detection, our CO limit
requires cloud tops at or above 3.3 mbar, and these clouds must be
opaque at a wavelength of 2 $\mu$m.  High clouds comprised of
submicron-sized particles are already present in some models, but may
not provide sufficient opacity to account for our CO result.  Cooler
model atmospheres, having smaller atmospheric scale heights and lower
CO mixing ratios, may alleviate this problem to some extent.  However,
even models $500$K cooler that the \citeauthor{sud03} model require
clouds above the 100 mbar level to be consistent with our
observations. Our null result therefore requires that
clouds exist at an observable level in the atmosphere of HD\,209458b,
unless this planet is dramatically colder than current belief.

\end{abstract}


\section{Introduction}

Doppler surveys have discovered more than 130 extrasolar planets
orbiting stars of near-solar type \citep{marcy03}.  About 15\% of
these extrasolar planets orbit within $\sim 0.05$ AU of their stars,
the so-called `hot Jupiters'. The small orbital radii of hot Jupiters
implies that they have significant probabilities ($\sim 0.1$) of
transiting their stars, and six transiting systems are known
\citep{charb00, henry, konacki03, konacki04, bouchy, alonso, pont}.
Transits provide opportunities to perform absorption spectroscopy of
the atmospheres of extrasolar planets \citep{ss00, hubbard, brown01}.
The extrasolar planet HD\,209458b is illuminated during transit by a
relatively bright star (V=7.65), permitting spectroscopy to high
signal-to-noise ratio (SNR), at high spectral resolution.
\citet{charb02} used transmission spectroscopy from the Hubble Space
Telescope Imaging Spectrograph to detect atomic sodium in the
atmosphere of HD\,209458b.  They discovered that sodium absorption is
weak relative to predictions \citep{ss00, brown01}, and several models
for this weakness have been suggested (reviewed by \citealp{seager03},
also see \citealp{fortney} and \citealp{burrows04}). The models that
explain the weakness of the observed sodium fall into two categories:
1) those invoking unique properties of sodium, such as a low sodium
abundance, or 2) those invoking some more general property of the
atmosphere, such as high clouds.  By searching for other spectral
features during transit, we can begin to discriminate between these
two classes of models.

In addition to sodium, atomic hydrogen has been detected in UV
absorption \citep{vidal03}, with evidence for oxygen and carbon
absorption ($< 3\sigma$ detections, \citealp{vidal04}).  Unlike sodium
in the hydrostatic atmosphere, the UV absorptions are relatively
strong ($5$--$15\%$) and extend to several planetary radii.
\citet{vidal03} conclude that they have observed an escaping planetary
coma.

All spectroscopic detections of extrasolar planetary atmospheres to
date have been made from space-borne platforms. But it should be
possible to make such detections from the ground.  Molecular
transitions in hot Jupiters often coincide with telluric transmission
windows, either because of the large planetary Doppler shift
\citep{wiedemann} or negligible level populations in the corresponding
telluric lines \citep{harrington}.  Carbon monoxide is a prime target
for ground-based detection.  It is predicted to be abundant at typical
hot Jupiter temperatures \citep{bursharp} and has strong and discrete
transitions in a favorable window near 2~$\mu$m.  Moreover, CO should
not be significantly depleted by photochemistry \citep{liang}, but
might actually be increased in abundance by mixing from greater depth
\citep{noll}.  However, like sodium, CO may be hidden by cloud
opacity. An infrared spectroscopic study of the 2~$\mu$m continuum by
\citet{rich03b} suggests the presence of substantial cloud opacity,
even on the stellar-facing hemisphere of the planet.

\citet{brown02} reported an exploratory attempt to detect CO in
HD\,209458b during transit using data from the Near-Infrared
Spectrometer (NIRSPEC) on the Keck~II telescope. They observed only a
single transit, and their spectra were degraded by poor weather, light
losses in the adaptive optics system, and prominent fringing in
NIRSPEC.  Nevertheless, they demonstrated sensitivity within a
factor of three of that needed to detect planetary CO in a plausible
model.  In this paper we report follow-up observations with NIRSPEC on
Keck~II that improve sensitivity by about a factor of ten.

\section{Observations}

We observed 3 transits of HD\,209458b on 19~August, 26~August, and
2~September 2002 UT using NIRSPEC \citep{mclean} on the Keck~II
telescope without adaptive optics.  Each night's observations consist
of a nearly continuous sequence of (typically) 45-sec integrations on
HD\,209458 from several hours before the center of transit to several
hours after.  We observed at wavelength $\lambda \sim\ $ 2.3~$\mu$m (K
band) with spectral resolution $\lambda/\delta \lambda = 25,000$,
where $\delta \lambda$ is the full-width at half-maximum (FWHM) of the
instrument profile and covers $\sim$ 3 pixels.  In order to facilitate
the greatest possible stability in the spectra, and the most efficient
temporal duty cycle, we did not nod the telescope.  We kept the star
at a single position on the $0.43 \times 24$ arc-sec slit.  Dark frame
exposure times bracket the stellar exposures and flat fields come from
spectra of a continuum lamp.  Each spectral frame consists of 6
non-contiguous echelle orders (3 of which contain CO lines).  Each
order covers $\sim 60\ $ cm$^{-1}$, ($\sim 0.06\ $cm$^{-1}$ per
pixel).  We used an order-sorting filter with 50\% transmission at
5400 and 3800 cm$^{-1}$ (1.84 \& 2.63 $\mu$m).

Passing cirrus clouds interrupted observations on 19~August,
especially in the first half of the night.  On 19~August and
especially early on 26~August, the grating position made uncontrolled
moves of a few pixels, sufficient to greatly complicate our analysis.
These uncontrolled grating moves are a recognized problem with NIRSPEC.  The
recommended solution (at that time) was to turn off the image
rotation.  Accordingly, we observed for most of 26~August and all of
2~September without using image rotation.

We recorded a total of 379, 456, and 354 spectra on 19~August,
26~August, and 2~September, respectively.  We reject 32 spectra from
19~August due to low signal level from cirrus cloud absorption, and we
reject the first 80 spectra taken on 26 August, due to the
uncontrolled grating motion (see Sec.~4).  The analysis of this paper
is therefore based on a total of 1077 spectra, each comprising $\sim
10^3$ independent spectral resolution elements.

\section{Theoretical Transit Spectra}

Searching for CO absorption in the transiting planet requires
averaging over multiple lines in several vibration-rotation bands to
achieve the required SNR.  We perform this averaging, in effect, by
fitting the observed spectral residuals (Sec.~4.2) to a theoretical
template spectrum using linear least squares (Sec.~4.6).  For
diagnostic purposes, we repeat this fit over a grid of theoretical
template spectra.  Our intent in this paper is to define the range of
atmospheric parameters allowed by our observations, not to construct
physically self-consistent models.  We therefore begin with a
radiative equlibrium temperature profile from \citet{sud03}, and
perturb it in an {\it ad hoc} fashion.  We represent clouds by setting
the total opacity to be infinite for pressures greater than either 1,
2, 3, 10, 37, 100, or 1000 mbar.  We also add various $\delta T$
values to the temperature at all pressures, choosing from $\delta T=$
 +200, 0, -200, -300, -500 and -700 Kelvins.  We define our fiducial
model to be the unperturbed \citet{sud03} T-P profile (${\delta}T=0$),
with cloud tops at $0.1/e$ bar $= 37$ mbar.  This cloud top pressure
was used as a fiducial model by \cite{charb02}, so this choice allows
us to judge the strength of any CO absorption relative to the sodium
detection.

Our computation of the theoretical template spectra utilizes the IDL
code developed by \citet{rich03b}, modified for tangent-path spherical
geometry.  After adding the $\delta T$ to the \citet{sud03} T-P
profile, we define the height scale ($h$) by integrating the equation
of hydrostatic equilibrium with surface gravity $g=945$ cm sec$^{-2}$
\citep{hst01}.  We choose the constant of integration so that $h=0$
occurs where the continuum optical depth in the tangent-path geometry
equals unity.  We take the planet's radius to be the distance from the
planet's center to the $h=0$ point.  In practice, the continuum
optical depth becomes unity very close to the pressure level assumed
for the cloud tops (see below).  Therefore, in the absence of stellar
limb darkening, the computed eclipse depth in the 2~$\mu$m continuum
will be $(R_p/R_s)^2 = 0.0146$ independent of the cloud top pressure
level, where $R_p$ is the planet radius, and $R_s$ is the stellar
radius, both from \citet{hst01}.

At each wavelength we compute the total tangent path optical depth as
a function of tangent height, using spherical geometry.  Refraction in
the planet's atmosphere can be significant in other transit-related
applications \citep{hui}, but will not have a measurable effect on our
transmission spectra, especially considering the much smaller
refractivity values at this long wavelength.  We therefore neglect
refraction in our computed spectra. The area of an atmospheric annulus
at each tangent height is weighted by the factor $1-e^{-\tau}$, and
the annuli are summed to give the total blocking area of the planet as
a function of wavelength.  The continuum optical depth is due to
collision-induced H$_2$-H$_2$ and H$_2$-He opacity \citep{borysow02,
jorg}.  The line opacity at each wavelength is computed by summing
opacities from all lines within some adjustable interval (typically 3
cm$^{-1}$), with the opacity profile of each line computed exactly
from the Voigt function.  We use a single damping coefficient for all
lines of a given species, based on averaging HITRAN 2000 data.  The CO line
wavelengths and strengths were taken from \citet{goorvitch}, including
isotopic lines in solar abundance ratios.  The ratio of the planet's
blocking area to the area of the stellar disk, with a correction for
stellar limb darkening, is the transit depth at that wavelength,
$d_{\lambda}$.

Our CO search requires that we consider the effect of water
absorption, because significant water lines overlay CO at some
wavelengths.  The water molecule exhibits millions of transitions in
the wavelength regions we have observed, most of these being quite
weak.  We evaluated the possible effect of a quasi-continuous opacity
due to numerous overlapping weak water lines, using the water line
data from \citet{partschwen}.  These data show a line density of
$\sim$~3000 lines per cm$^{-1}$ in the 2~$\mu$m region, potentially
sufficient to obscure the contrast in the CO transit spectrum.  We
scaled the strengths of all the water lines to the $\sim$ constant
temperature above the cloud tops, and summed their scaled strengths in
bins of 0.01~cm$^{-1}$ width.  The total strength in each bin was then
treated as a single pseudo-line, with lower state energy of zero.
This treatment will be relatively accurate for the upper layers of
HD\,209458b's atmosphere, because our bin size is comparable to the
Doppler width, and because the temperature in the \citet{sud03} model
remains relatively constant at the greatest heights.  The thermochemical
equilibrium abundances of CO and water were computed from the formulae
given by \citet{bursharp}.  Since water is likely to be depleted by
photolysis at high altitudes \citep{liang}, our assumption of
equilibrium water abundance is conservative for the purpose of
detecting CO.

Our tangent-path spectrum code does not attempt to modify the
pressure-temperature relation from the stellar-facing to anti-stellar
hemispheres.  However, we include template spectra where we set the CO
mixing ratio to zero on the anti-stellar side, simulating the effect
of a large temperature difference \citep{showman}.  We use a constant
radial velocity along each line of sight.  However, we do include the
effect of planetary rotation, assumed to be tidal-locked solid-body
rotation. We shift the spectrum in velocity corresponding to multiple
azimuth points in the atmospheric annulus, and add these shifted
spectra to produce a rotationally-broadened template.

Our tangent-path code has been checked in two ways.  First, its
ancestor code (for normal incidence spectra) was checked against
independent calculations by Sara~Seager \citep{rich03b}, and second,
we compared the present calculations against selected individual CO
lines in the transit spectra computed by \citet{brown01}.  Figure~1
shows our template spectra, based on convolving the transit depths
($d_{\lambda}$) to the spectral resolution of NIRSPEC. The difference
in transit depth between the continuum and the strongest CO lines is
$\sim 0.001$ in the unconvolved (0.5 km sec$^{-1}$ resolution)
spectra.  This reduces to $\sim 0.0004$ at NIRSPEC resolution.

\section{Data Analysis}

A first analysis of the data for the best night (26~August) was done
(by TMB) using the cross-correlation/singular-value-decomposition
method described by \citet{brown02}.  From this first look, TMB
concluded that no CO absorption could be detected in the spectra, to a
limit $\sim 3$ times less than the expected amount.  It was also
evident that the data were of sufficient quality that the
singular-value-decomposition formalism was no longer necessary.  We
have accordingly analyzed the full set of data using more conventional
methods.  Our primary analysis uses linear least squares to correlate
the data against the ensemble of lines in each template spectrum.  We have
also checked the least squares results using cross-correlation
calculations.

The data analysis consists of: 1) extraction of spectra from the raw
spectral frames, 2) wavelength calibration and correction for telluric
absorption to yield spectral residuals, 3) injection of fake signals
into copies of the extracted spectra, to be analyzed in parallel with
the real data, 4) evaluation of the noise level as a function of time,
wavelength, and temporal frequency, 5) fitting the template spectrum
to the residuals by linear least squares to estimate the amplitude of
CO absorption in each spectrum, and 6) fitting a transit curve to the
CO absorption amplitudes.

\subsection{Spectrum Extractions}

Our choice to observe without nodding places stringent demands on
the spectrum extraction procedure, because effects normally removed by
nodding (sky emission and some artifacts such as `hot pixels') have to
be removed by numerical filters. We therefore performed spectrum 
extractions using a custom IDL code, not the NIRSPEC facility codes,
and we describe this process in more detail than usual.

For each spectral frame, we subtract a dark frame and divide by a
flat-field frame. The flat-field frame is a pixel-by-pixel median of a
series of continuum lamp exposures, each normalized to the same total
intensity.  For each spectral frame, we compute a pixel-by-pixel
median of five frames centered on the given frame. We subtract the
median frame from the given frame to form a difference frame.  A
two-stage, 1D median filter is applied to rows (along the dispersion
direction) of the difference frame, with decreasing width and
threshold.  This removes energetic particle events and the most
obvious hot pixels.  In addition, we compute the pixel-by-pixel
variance in the series of flat-field (continuum lamp) spectra, and we
use these variances with an appropriate threshold to replace noisy
pixels (`warm pixels') in the difference frame. The filtered
difference frame is then added back to the median frame to produce a
cleaned version of the given frame.

The curvature of each echelle order on the cleaned frame is determined
by fitting a Gaussian to the spatial profile of the star along the
slit at each wavelength.  A 4th order polynominal fit to the Gaussian
centroid positions defines the spatial center of the order as a
slowly-varying function of wavelength.  The spatial profile along the
slit at each wavelength is then spline-interpolated to a common
spatial grid to define the average spatial profile for each echelle
order. This spatial profile is computed separately for each echelle
order, but we found that it does not vary significantly with
wavelength in a given order.  The spatial profile is fit to the
original data at each wavelength by linear least squares, with free
parameters being amplitude and zero-point in intensity.  The fitted 
amplitude is equivalent to the optimal spectrum value at that
wavelength \citep{horne}, and the background subtraction occurs
automatically via the zero-point of the fit.  In this process, we
reject further warm pixels based on the residuals from the fit of the
spatial profile to the data.  Our procedure ignores the tilt of the
slit.  In principle this can degrade the spectral resolution, but
degradation in resolution is negligible for our data because each
spectrum is narrow in spatial extent ($\sim 5$ pixels~FWHM), and
because the CO absorption fortunately occurs in echelle orders where
the tilt of the slit is minimal ($\leq 0.04$ pixels in $\lambda$, per
spatial pixel). We verified the resolution of the extracted spectra by
fitting Gaussian profiles to determine the widths of minimally-blended
telluric lines, and we obtained good agreement with the nominal
NIRSPEC resolution.

\subsection{Telluric Corrections and Wavelength Calibration}

The extracted spectra are generally not coincident in wavelength. We
therefore shift the spectra to make them coincident and apply a
wavelength calibration to the average spectrum as follows.  We begin
by computing an order-by-order average of the non-coincident spectra.
Each order of a given spectrum is then spline-shifted in small
increments (0.025 pixels) and fitted to the average spectrum at each
shift value by stretching in intensity using linear least squares.
The rms difference in intensity is tabulated {\it vs.} shift, and a
parabolic fit finds the `best' shift from the minimum rms
difference. Normalization in intensity uses a continuum fit.  About 10
continuum windows are selected in each order, and the continuum is
derived from a 4th order polynominal fit to the intensity of each
spectrum in those windows.  We divide the observed spectrum by the
{\it wavelength-integrated} intensity under the continuum fit (not a
point-by-point division).  We then fit the log of intensity to airmass
at each wavelength using linear least squares and correct each
spectrum, at each wavelength, to an airmass of unity.  (For 19~August
we make separate airmass corrections for the first and second halves
of the night, since sky conditions changed.)  The airmass corrections
ranged in value from $\sim 0.03$ magnitudes per airmass for the best
windows to $> 1$ in telluric lines.

If telluric absorption were strictly proportional to airmass, the
spectra corrected as described above would vary only due to absorption
by the planet during transit.  However, Fourier analysis of the
intensities at each wavelength show power spectra whose amplitudes
increase sharply for time intervals greater than $\sim$ 15 minutes (2
cycles per hour).  We denote this as `1/f' noise (without implying
that the noise power is strictly proportional to the inverse of
frequency).  The 1/f noise is more pronounced near telluric water
lines, but is evident to some degree virtually everywhere in the
spectra.  These ubiquitous slow variations in the terrestrial
atmosphere cannot be completely corrected using any proportional-to-airmass
technique, and they are the limiting source of error in our analysis.
Fortunately, they are partially correlated between closely spaced
wavelengths, so we can remove them to some degree using a high-pass
digital filter.  For a given wavelength $\lambda$, we compute the
average out-of-transit spectrum over the wavelength range $\lambda -
\delta\lambda$ to $\lambda + \delta\lambda$. where $\delta\lambda$ is 
3~times the NIRSPEC spectral resolution.  We fit this short fiducial
spectral segment to the corresponding portion of each individual
spectrum, scaling it in intensity by using linear least squares.  The
difference between the best scaled fiducial segment and the real
spectrum at wavelength $\lambda$ and time $t$ is denoted as the
residual intensity $r_{\lambda t}$, in units of the stellar continuum.
The $r_{\lambda t}$ are the fundamental data in which we search for
real planetary CO absorption during transit.  The computation and
fitting of the fiducial segment is repeated at each wavelength.  The
choice of $\delta\lambda$ equal to 3 NIRSPEC resolution elements (9
pixels) is optimal for removing telluric effects while passing real
signal. This optimal value is selected by trial and error: we vary
$\delta\lambda$ and monitor the magnitude of the remaining low
frequency telluric variations, and the amplitude of the transit in
data containing a fake transit signal.

Wavelength calibration is performed, order-by-order, using the average
of the coincident and normalized spectra. (We do the actual
calibration in frequency, not wavelength.)  In each echelle order we
choose about 12 minimally-blended telluric calibrating lines, and we
fit parabolas to their profiles within one spectral resolution element
of the intensity minimum. These fits to the line cores minimize
contamination by line blends and define the precise pixel positions of
the telluric calibrating lines.  To relate the pixel positions to
wavenumber, we use the digital version of the high-resolution solar
spectral atlas \citep{lw91}. Convolving this atlas to NIRSPEC
resolution, we perform the same parabolic fits as on the NIRSPEC
lines.  This yields effective wavenumbers for the telluric calibrating
lines at NIRSPEC resolution. A quadratic fit of the pixel positions to
the effective wavenumbers provides coefficients of the wavelength
calibration.  One can judge the precision of this calibration from the
scatter in the fits of wavenumber to pixel position, which is 0.01 to
0.02~cm$^{-1}$ (0.18 to 0.35 pixels), depending on the echelle order.
Figure~2 shows the wavelength-calibrated spectrum in echelle order 33,
for 26~August 2002 (compare to Figure~1 of \citealp{brown02}).  All of
the structure visible in Figure~2 is either telluric absorption
(primarily methane), or instrumental effects (fringing in NIRSPEC).
The fringing seen in our spectra is similar in period and amplitude to
that seen by \citet{brown02}, but is more stable.

\subsection{Noise Level and Fringe Removal}

The noise level $\sigma_{\lambda t}$ is a function of time,
wavelength, and temporal frequency.  Time dependence derives from the
variation of signal strength with airmass, and general sky clarity,
and wavelength dependence derives from the grating blaze function and
the telluric line spectrum.  The dependence on temporal frequency
derives from the 1/f noise described above.  At a given wavelength, we
compute the variance of the $r_{\lambda t}$ residuals in two
bandwidths, with time varying, by summing the power spectrum over the
appropriate limits.  We denote the standard deviations as
$\sigma_{\lambda}^h$ and $\sigma_{\lambda}^l$, where $h$ and $l$
denotes frequencies higher and lower than 2 cycles per hour,
respectively.  The square root of the quadratic sum of
$\sigma_{\lambda}^h$ and $\sigma_{\lambda}^l$ is denoted
$\sigma_{\lambda}$, the total noise level in intensity at wavelength
$\lambda$.  $\sigma_{\lambda}$ is shown {\it vs.} wavelength in Figure~3
for echelle order 33, on 26~August, 2002.  The best wavelengths
exhibit SNR ($=1/\sigma_{\lambda}) \sim$~300, but the noise level
increases sharply near some telluric lines.  We therefore reject some
wavelengths from our analysis, based on the $\sigma_{\lambda}$ values.
Typically we mask out those wavelengths where $\sigma_{\lambda}$
exceeds 200\% of a `baseline' noise level, obtained as the lower
envelope of the $\sigma_{\lambda}$ values.  The masked wavelengths are
overplotted with diamond symbols on Figure~3, and they invariably
correspond to wavelengths near telluric lines (compare
Figures~2~\&~3).

In addition to the $\sigma_{\lambda}$ values, we also have an estimate
of how the noise varies from spectrum to spectrum, from the fits of
individual spectra to the average spectrum, a by-product of making the
spectra coincident in wavelength.  We denote the average noise level
of the spectrum at time $t$ as $\sigma_{t}$. We can therefore estimate
the noise level expected for each pixel in each spectrum, in each
bandwidth, $\sigma_{\lambda t}^h$, and $\sigma_{\lambda t}^l$, and the
total over all frequencies $\sigma_{\lambda t}$, by assuming that the time
and bandwidth variations are independent of $\lambda$ (an
approximation), and requiring that the sum of $\sigma_{\lambda t}^{2}$
equal $N \sigma_{\lambda}^{2}$ for each $\lambda$, in each bandwidth,
where $N$ is the number of spectra for that night.  The best
$\sigma_{\lambda t}$ values are lower than the $\sigma_{\lambda}$
values shown in Figure~3, because $\sigma_{\lambda}$ includes spectra
taken over all airmass values. Our best spectra have peak signal
levels per column of $\sim 1.4 \times 10^{5}$ electrons, so we should
reach the Poisson electron statistical limit (SNR~$\sim$~380) for the
best $\sigma_{\lambda t}$.  The $\sigma_{\lambda t}$ distribution has
63\% of its SNR values between 200 and 400, with 26\% having SNR
$<200$, and 11\% with SNR $>400$.  This sharper cutoff at high SNR is
consistent with reaching the photon noise limit.

There are two additional error sources in our spectra.  First, there
is the optical fringing with two distinct periods, noted by
\citet{brown02}. Both fringe systems are found in our data; however,
they are less time-variable than experienced by \citet{brown02}. Some
of this optical fringing is removed by the digital filter which
produces the $r_{\lambda t}$.  However, Fourier analysis of the
$r_{\lambda t}$ data reveals that both fringe systems remain
detectable.  When the residual fringes' amplitudes exceed 0.25 in the
log of the average power spectrum per echelle order per night, Fourier
notch filters remove them.  Second, extra noise is ubiquitous near the
Nyquist frequency (0.5 cycles per pixel) in the power spectra of the
$r_{\lambda t}$ with $\lambda$ varying.  This probably derives from
the array detector electronics.  We remove it prior to the digital
filter step by zeroing the Nyquist frequency in the Fourier spectrum.
Figure~4 presents the stacked, wavelength-coincident spectra and the
$r_{\lambda t}$ values for echelle order 33 on 26 August 2002, as
false color images.

A complete understanding of the errors requires comparing the
distribution of $r_{\lambda t}$ to a normal error curve, and also
examining whether the noise power per unit bandwidth is constant (white
noise).  Although the error level varies with time and wavelength, the
ratio of the individual residual values to their expected error,
$\frac{r_{\lambda t}}{\sigma_{\lambda t}}$, should approach a Gaussian
error distribution, with a standard deviation of unity.  Figure~5
(upper panel) shows this distribution totaled over all 3 nights and
all 3 echelle orders, on a log scale.  Within $3\sigma$ this
distribution is very close to the Gaussian error curve.  However, the
data exhibit an excess of points in the wings beyond $\pm 3\sigma$.
Specifically, 0.5\% of the $3.02 \times 10^{6}$ points lie beyond
$3\sigma$, and 0.05\% are beyond $4\sigma$. A Gaussian distribution
predicts 0.27\% and 0.006\% beyond $3\sigma$ and $4\sigma$
respectively.  Using a Monte-Carlo simulation, we verified that the
number of excess points in the wings of the data distribution are not
sufficient to have a significant impact on our results, nor to affect
our error analysis significantly, particularly since they are
uncorrelated with transit phase.  A more serious problem is the fact
that the noise power per unit bandwidth is not constant, i.e. the 1/f
character previously mentioned. We compute the power spectrum of each
$r_{\lambda t}$ series, with $t$ varying, using a Lomb-Scargle
algorithm \citep{press}.  The average of these power spectra for all
$\lambda$ over all three nights is shown as the lower panel of
Figure~5.  The sharp increase in noise power below $\sim$~2 cycles per
night is quite evident; this increased noise at low frequencies would
be about 2 times worse (0.3 in the log) without the digital filter
technique described above.

\subsection{Validation Using Fake Signals}

In order to verify that our analysis is sensitive to the low levels of
absorption expected in transit spectra, we have injected fake signals
into the data and recovered them with our analysis procedures.  The
first fake signal is generated from the CO template spectrum of the
fiducial model (Figure~1, solid line), Doppler shifted into the frame
of the planet, and convolved to NIRSPEC spectral resolution.  We
adopted a heliocentric radial velocity for the system of $V_r =
-14.76$ km sec$^{-1}$ \citep{nidever}, and a planetary orbital
velocity amplitude of 142~km~sec$^{-1}$.  The orbital period and
transit times are very precisely known \citep{wittenmeyer, schultz,
charb03}, and we corrected for the Earth's orbital motion, both in
radial velocity and light travel time. A second fake signal is
generated using the CO+water template spectrum (Figure~1, dotted
line), and similarly transformed to the frame of the planet.  We
injected the fake signals into the data after the wavelength
calibration, but before the digital filter and fringe removal. We
therefore analyze three sets of data in parallel: the real data, and
two copies of the real data with injected fake signals.  Each of these
data sets is analyzed against a grid of template spectra, where we
offset the temperature in the model atmosphere, and vary the cloud
height.  In some cases we also set the CO mixing ratio to zero on the
anti-stellar hemisphere of the planet, to simulate the worst-case
effect of a large diurnal temperature difference.

\subsection{The Stellar Spectrum}

Our spectra contain significant information concerning the spectrum of
the star, i.e. HD\,209458a.  Two aspects of the stellar spectrum are
relevant to the transit analysis.  First, we have searched for stellar
CO absorption.  In the solar spectrum, the first overtone CO features
are relatively prominent, having line depths, when convolved to
NIRSPEC resolution, of $\sim 6$\%.  Stellar lines are removed by the
procedures that compute the $r_{\lambda t}$ residuals, so we have to
identify them prior to removal of telluric lines. Also, stellar lines
are not significantly modulated by the transit. Hence, they are
somewhat difficult to detect.  We therefore overlaid and averaged 26
lines of stellar CO lying in good telluric transmission windows.  We
see no evidence of a CO signature in the stacked spectra, with the
confusion limit being determined by the telluric background at the
$\sigma \sim 0.5$\% level.  CO lines of solar strength would be easily
seen ($\sim 12\sigma$), but HD\,209458a shows no detectable CO
features. We considered whether the system might be lacking in carbon
and/or oxygen, but the star has normal metallicity \citep{mazeh}, and
2- to 3$\sigma$ evidence for both carbon and oxygen is seen in the planet's
UV transit spectrum \citep{vidal04}.  It is more likely that the
non-appearance of stellar CO is related to the temperature of
HD\,209458a, which is slightly hotter than the sun \citep{mazeh,
ribas}.

Although stellar CO is not detected in our spectra, we do detect the
few stellar atomic lines that are expected.  We used 5 atomic lines
to derive the star's radial velocity.  We identified the lines
using the \citet{lw91} solar atlas, and fit parabolas to their line
cores in both our Keck spectra and the solar atlas convolved to
NIRSPEC resolution.  On this basis we derive a heliocentric radial
velocity for the system of $-15.4 \pm 0.5$ km sec$^{-1}$, agreeing
with the more precise \citet{nidever} value, within our error.  This
serves as an independent check on our velocity transformation between
the planet frame and the heliocentric frame.
  
\subsection{Transit Spectrum Amplitudes and Transit Curve Fit}

For each observed spectrum (at time $t$), we determine the degree to
which the template transit spectrum ($d_{\lambda}$) is present in the
$r_{\lambda t}$ values.  We Doppler shift the template spectrum to the
frame of the planet, convolve it to NIRSPEC resolution (see above),
and filter it exactly as we filter the data.  This transformed and
filtered template spectrum ($d^{\star}_{\lambda}$) is used as the
independent variable, with the $r_{\lambda t}$ values as the dependent
variable, in a linear regression at each $t$:

\begin{equation}
r_{\lambda t} = z + a(t)d^{\star}_{\lambda},
\end{equation}

where $z$ is a zero-point constant having no significance, and $a(t)$
is the statistical best-estimate of the amplitude of the template
transit spectrum in the $r_{\lambda t}$ at that time (this treatment
follows \citealp{rich03a}).  If the template spectrum is an exact
description of the planet's transit spectrum, then the $a(t)$ will lie
on a transit curve which is zero out of transit and dips to -1 during
transit (the negative sign denotes absorption, the $d_{\lambda}$ being
proportional to the blocking area). As an added dimension to this
calculation, we vary the assumed heliocentric radial velocity of the
HD\,209458 center of mass, by up to $\pm 235$ km sec$^{-1}$ ($\pm 50$
pixels), and we tabulate $a(t)$ for each $V_r$.  Since line structure
in the template corresponds to entirely different pixels in the data
as $V_r$ is varied significantly, this added dimension serves to check
the errors and significance of our results.

For the 26~August data, our first 80 spectra showed particularly
troublesome uncontrolled grating motion.  We found a systematic error in
results based on these data, evident as a correlation between the
spectrum shift and the $a(t)$.  These 80 spectra were all
taken prior to the start of transit.  Since this portion of the
transit curve is adequately sampled by other nights, we have deleted
the first 80 spectra taken on 26~August from our analysis, as noted in
Sec.~2. 

We input the $\sigma_{\lambda t}$, $\sigma_{\lambda t}^l$, and 
$\sigma_{\lambda t}^h$ values to the least squares fits in order to
estimate the errors in the $a(t)$ over the corresponding
bandwidths, denoted ${\sigma}_a$, ${\sigma}_a^l$, and ${\sigma}_a^h$. The
${\sigma}_a$ are typically as large, or larger, than the amplitude
itself (SNR$\lesssim$~1).  Nevertheless, the linearity of the analysis
preserves the fidelity of the signal when the amplitudes from 1077
spectra are considered, and the expected SNR on the transit curve
becomes sufficient to expect detection of CO absorption.

Fitting a theoretical transit curve to the $a(t)$ {\it vs.} time
implicitly involves averaging them.  This requires proper accounting
for the errors as a function of frequency.  For example, spectra taken
in quick succession will produce an average $a(t)$ whose high
frequency error of the mean decreases as the inverse square root of
their number.  But averaging spectra over brief intervals will not
similarly reduce the low frequency errors, which can be correlated
over short times.  We are not aware of any rigorous methodology for
treating non-white quasi-Gaussian errors in data analysis.  We
therefore implement the approach of approximating the full Fourier
spectrum by two frequency bands, above and below 2 cycles per hour.
We average the $a(t)$ for each night in 15-minute bins (0.003 in
phase).  Within a given bin on a given night, the high frequency
errors apply, and the low frequency errors apply when comparing one
bin to another.  The binning involves a weighted average: within bin
$i$ on night $j$ the $a(t)$ are weighted by the inverse square of
their ${\sigma}_a^h$.  The error associated with the average of bin
$i$ on night $j$ is given as:

\begin{equation}
\sigma_{ij} = ( \sum ({\sigma}_a^h)^2/N^2 + ({\sigma}_a^l)^2)^{\frac{1}{2}},
\end{equation}

where the sum extends over the $N$ amplitudes in bin $ij$ and
${\sigma}_a^l$ is the typical low frequency error common to all points
in the bin (the low frequency errors do not vary strongly within each
$ij$ bin).  Eq.(2) is easily derived from first principles, and we
have verified it using numerical simulations.  A grand average
amplitude at phase $i$ results from combining bins at all $j$ (all
nights), weighting the $ij$ bin averages by the inverse square of
$\sigma_{ij}$.  We similarly combine the $\sigma_{ij}$ to yield the
error for the grand average amplitude at phase $i$.

We fit a theoretical transit curve to the grand average amplitudes and
their errors. The shape of the theoretical transit curve is generated by
numerically simulating the passage of the planet across a
limb-darkened star, and the depth is normalized to unity, to be
consistent with the amplitude retrievals (see above).  The stellar and
planet parameters were adopted from \citet{hst01}, and the stellar
limb-darkening at 2~$\mu$m was taken from the solar observations of
\citet{pierce}.

\section{Results}

Figure~6 plots the retrieved CO amplitudes, $a(t)$, for the fiducial
model {\em vs.} phase.  The 1077 individual $a(t)$ are shown in the
top panel, with error bars supressed for clarity, but with different
colors used for the three nights. The middle panel shows the grand
average amplitudes binned over phase (bin width =0.003), with error
bars added.  To clarify our sign conventions, recall that points with
negative amplitudes denote absorption. We fit a transit curve dipping
to negative values during transit, so the fit of a negative curve to
negative CO amplitudes should retrieve a positive depth (+1.0 if the
fiducial model is correct).  The transit curve is fit to the binned
amplitudes in the middle panel, giving a depth of $-0.09 \pm 0.14$ in
model units, i.e. no evidence of CO absorption matching our template
spectrum during transit. The solid line shows this best fit. The
square root of the reduced chi-squared of the fit, $\sqrt{\chi^{2}/M}
= 1.19$, where $M$ is the number of degrees of freedom. Hence, the
scatter in the binned amplitudes is somewhat greater than the error
bars, which result from propagating the error in the $r_{\lambda t}$
values consistently through our linear analysis. The dashed red line
in the middle panel shows a transit amplitude of unity, which is
obviously incompatible with the data.

The lower panel of Figure~6 shows the binned fake data with transit
curve fit.  Our analysis recovers 82\% of the input signal, with a
least squares best fit transit depth of $0.82 \pm 0.14$; this transit
is shown as the solid red line.  Note that only the relative depth of
the transit curve is significant in these fits: the zero point
displacement in amplitude units has no significance.

The 1/f noise in the $r_{\lambda t}$ produces a noticeable signature
on Figure 6. Slow variations are evident in the top panel, giving the
points in each color a quasi-patterned appearance.  These variations
ultimately derive from the variability of the terrestrial atmosphere.
To investigate the error and significance of our result further,
Figure~7 shows the depth of the best fit transit curve {\it vs.} the
heliocentric $V_r$ of the HD\,209458 center of mass (re-doing the
analysis of the $r_{\lambda t}$ for each assumed $V_r$).  The
dispersion of the waveform shown on Figure~7 should be the same as the
average error associated with the transit curve depth ($\sim 0.14$).
The Figure~7 dispersion is 0.22; if we use this more conservative
value as the error in transit curve depth, and considering the 82\%
efficiency of recovering the fake signal, the fiducial model is
rejected at $\sim 4\sigma$ significance.  The dashed line on Figure~7
shows the result from the fake CO signal, which is clearly detected at
the correct $V_r$ of the system.

Our grid of template spectra provides several ways to explore physical
reasons for the non-appearance of a CO transit.  For example, we
hypothesize a large temperature difference between the stellar-facing
and anti-stellar hemispheres of the planet, with CO being absent on
the anti-stellar side.  As the template lines weaken, the error level
increases, because the same data are being fit to smaller contrast in
the template.  When the error envelope, to some specified level of
significance, expands to include a transit depth of 0.82 (the fake
signal recovery level), that particular template spectrum can no
longer be rejected. Setting the CO mixing ratio to zero on the
anti-stellar side of the fiducial model produces a best fit transit
depth of $-0.11 \pm 0.16$, with the Figure~7 dispersion increasing to
0.25.  This case is rejected at greater than $3\sigma$ significance,
so removal of CO on the anti-stellar hemisphere has only a small
effect on our results.  The effect is small because the strong CO
lines in the template are saturated.

Other physical effects that could account for our result are high
clouds, or decreased temperature in the upper atmosphere of the
planet.  We have repeated our analysis over our entire grid of
template spectra.  None of the templates gives a significant detection
of a CO transit.  We compute the maximum cloud top pressure consistent
with this null detection for each $\delta T$, by interpolating to find
where the error envelope allows a transit depth of 0.82 (the fake
signal transit depth).  We take the size of the error envelope to be
three times the formal error in transit depth (as in Figure~6 fits),
or the maximum excursion in transit depth {\it vs.} $V_r$ (as in
Figure~7), whichever is greater.  The maximum cloud top pressure is
plotted {\it vs.} the $\delta T$ added to the temperature profile in
Figure~8. A separate relation is shown for the case when CO is absent
on the anti-stellar hemisphere.  Investigators who wish to make their
own evaluation of models should use the criterion that the strong 2-0
R-branch lines between 4320 and 4330 cm$^{-1}$ have depths during
transit less than 1.6 parts in $10^{4}$ in units of the stellar
continuum (3$\sigma$ limit), at a spectral resolving power of 25,000.
This observational criterion is weakly model-dependent because we
derive it using a template spectrum (Figure~1) wherein the relative
strengths of the CO lines vary slightly with temperature.

Water absorption has been included in one version of our model
template (Figure~1, dotted line), and carried through our analysis as
a second fake signal.  To verify that water does not mask the presence
of CO in our analysis, we first analyze the CO+water fake signal using
the template containing only CO. This analysis yields a transit depth
of $0.69 \pm 0.14$.  This is less than the $0.82$ depth recovered from
the CO-only fake signal, albeit the same within the errors.
Nevertheless, to guarantee maximum sensitivity to CO in the presence of
water lines, we have analyzed the data with a full grid of CO+water
templates.  The CO+water template in the fiducial model recovers the
fake CO+water signal with essentially the same efficiency (80\%) that the
CO-only template recovers the CO-only fake signal (82\%).  We have
verified that our results (i.e., Figure~8) do not change significantly
when the data are analyzed using the CO+water template grid.

\section{Discussion}

Our upper limit for CO is similar to the sodium detection
\citep{charb02}, in the sense that absorption from both Na and CO is
significantly weaker than predicted by a fiducial model with cloud
tops at 0.1/e bars (=37 mbar). Several published explanations for the
weakness of sodium invoke some relatively specific mechanism that
would not necessarily apply to CO (e.g., \citealp{barman}).  Our upper
limit is interesting because fewer explanations can account for the
weakness of both species.  One of these viable explanations is the
presence of high clouds. To be strictly consistent in comparing with
the \citet{charb02} interpretation of their sodium observations, we
must refer to the highest temperature on Figure~8, $\delta T =
+200$K. \citet{charb02} followed \citet{brown01} in using an early
version of the \citet{sud03} model that has $T=1150$K in the upper
layers, 200 Kelvins hotter than is the model published by
\citet{sud03}.  At $\delta T= 200$K, our results require a maximum
cloud top pressure of 3.3~mbar (Figure~8).  Using the \citet{sud03}
model without perturbation increases this to 5.7~mbar.  Clouds occur
at pressures as low as 1~mbar in some models.  Not only do they
reproduce the sodium result, they are helpful in accounting for the
relatively large radius of the planet \citep{fortney, burrows03}.
However, the high forsterite cloud discussed by \citet{fortney} is
comprised of small particles ($\leq 0.5~\mu$m), whose visible opacity
is dominated by scattering.  At the much longer wavelength of the CO
bands, scattering opacity by submicron particles will be greatly
reduced (an issue discused by \citealp{brown01}), and such clouds may
not suffice to explain our CO result.  Indeed, \citet{burrows04} show
a spectrum (their Figure~1) based on data from \citet{fortney}, that
illustrates weakened sodium, but still shows prominent CO features.
Cooler atmospheres would allow the IR-opaque cloud tops to occur at
higher pressures (Figure~8), and might prove easier to reconcile with
our results. Note that a cooler atmosphere for HD\,209458b could also
explain the non-detection of a 2~$\mu$m continuum peak by
\citet{rich03b}.

It is interesting to note that even very cool temperatures for
HD\,209458b still require opaque clouds at pressures much less (i.e.,
greater geometric heights) than the point where a cloudless atmosphere
becomes opaque in the tangent geometry.  For example, Figure~8
requires cloud tops above 100 mbar even for a model 500K cooler than
the \citet{sud03} model.  As we decrease the temperature in the
\citet{sud03} model, the mixing ratio of CO becomes negligible at the
greatest heights, but a significant transit signature should persist
from CO at higher pressures.  Our null result therefore requires that
clouds exist at an observable level in the atmosphere of HD\,209458b,
unless this planet is dramatically colder than current belief.

We have considered alternative interpretations of our result. Since
the observed spectral resolution is high, a stringent requirement is
that the line wavelengths be accurate.  Fortunately, errors in line
wavelengths are unlikely to be significant to our analysis.  We
inspected the much higher-resolution solar atlas \citep{lw91}, and
were able to confirm wavelengths even for lines of insignificant
importance to our template.  For example, in the solar spectrum we
find the 7-5~R(59) transition at 4083.797~cm$^{-1}$, {\it vs.}
4083.793~cm$^{-1}$ in our template \citep{goorvitch}.  We also
consider the possibility that the absorbing levels of CO are
depopulated high in the planetary atmosphere, i.e., not in local
thermodynamic equilibrium (LTE).  Note that departures from LTE are
only significant as they affect the lower state of the absorbing
transition; transit spectroscopy is insensitive to the upper state
population.  The strongest CO lines in our template are the P and R
branches of the 2-0 and 3-1 bands, so it suffices for the $v=1$ level
to be in LTE for our analysis to be valid.  \citet{ayres} discuss the
radiative and CO-H$_2$ collisional rates for $v=1$ to $v=0$
relaxation.  Using their Landau-Teller parameters for H$_2$
collisions, we calculate that the $v=1$ to $v=0$ collisional and
radiative rates become comparable at 1~mbar.  Therefore departures
from LTE cannot solely explain our result, which requires weakening of
CO at tens of mbar, where collisional rates dominate.

\acknowledgments We thank the Keck support staff, particularly Randy
Campbell and Grant Hill, for assistance with NIRSPEC, and we thank
Dave Sudarsky and Adam Burrows for providing the temperature/pressure
profile of HD\,209458b. Sara Seager made comments which improved this
paper significantly. Portions of this work were supported by NASA's
Origins of Solar Systems program.  The authors wish to recognize and
acknowledge the very significant cultural role and reverence that the
summit of Mauna Kea has always had within the indigenous Hawaiian
community.  We are most fortunate to have the opportunity to conduct
observations from this mountain.



\clearpage


\begin{figure}
\plotone{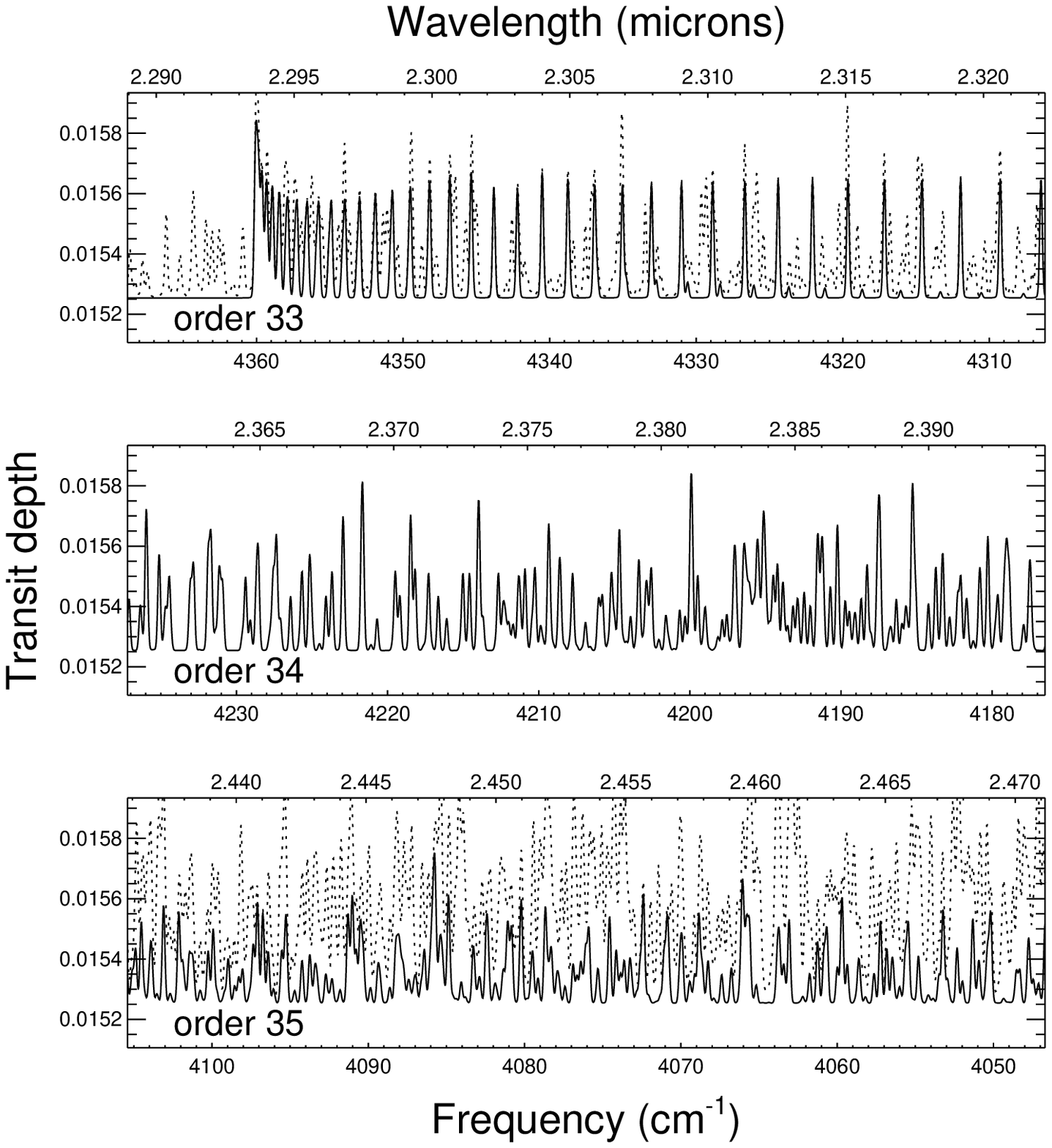}
\caption{Theoretical transit depth for our fiducial model (cloud tops
at 37 mbar) in the cases where we consider CO alone (solid line)
and CO plus water (dotted line).  The template spectra are plotted
over the range covered by the 3 NIRSPEC echelle orders that contain
CO, and they have been convolved to NIRSPEC resolution.
\label{fig1}}
\end{figure}

\begin{figure}
\plotone{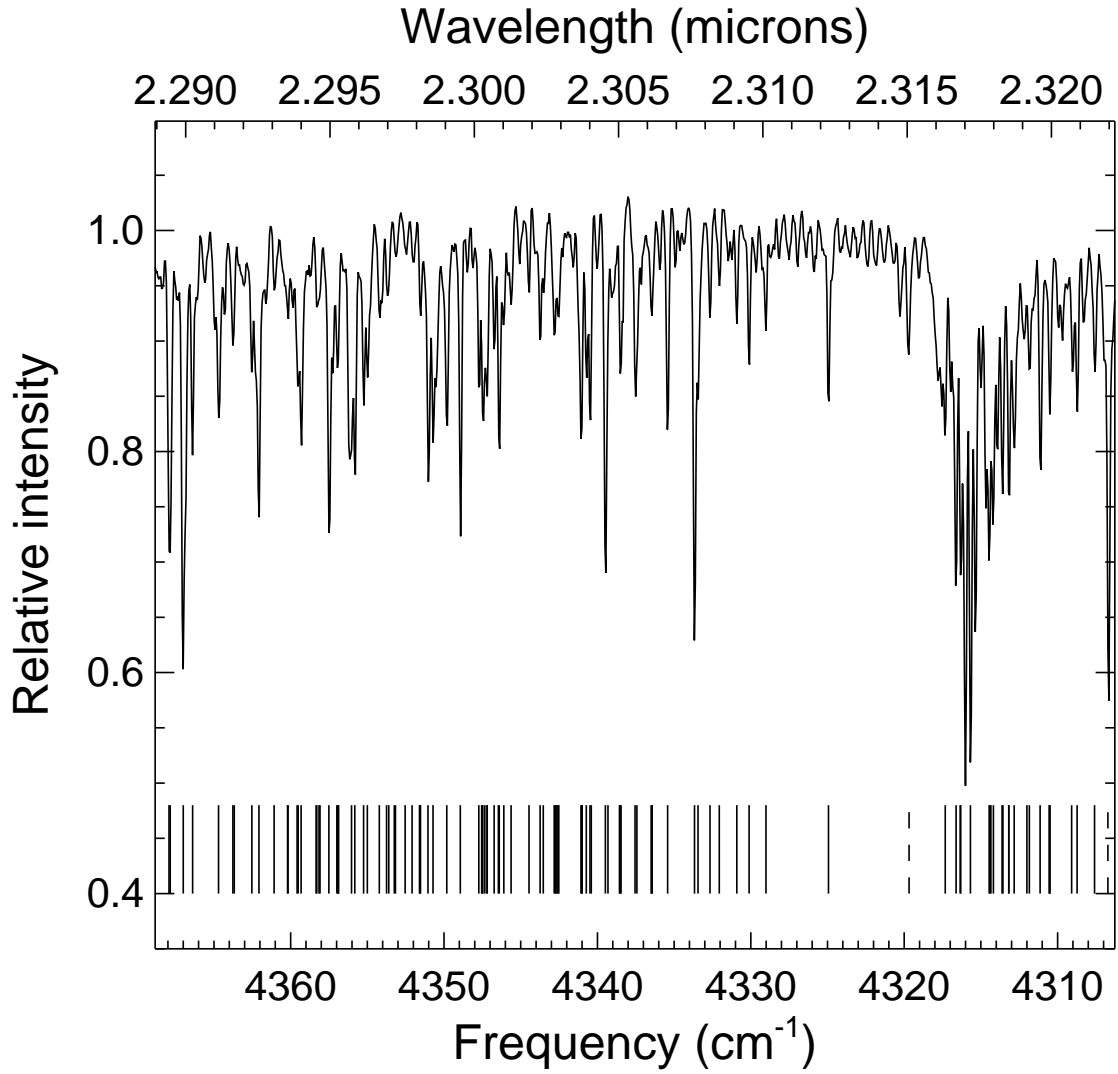}
\caption{Average spectrum of HD\,209458 in echelle order 33 on 26
August 2002.  Most of the spectral structure is due to telluric lines
of methane (solid tics) and water (dashed tics). Instrumental
fringing is particularly noticeable in the 4320-4328 cm$^{-1}$
region. \label{fig2}}
\end{figure}

\clearpage 

\begin{figure}
\plotone{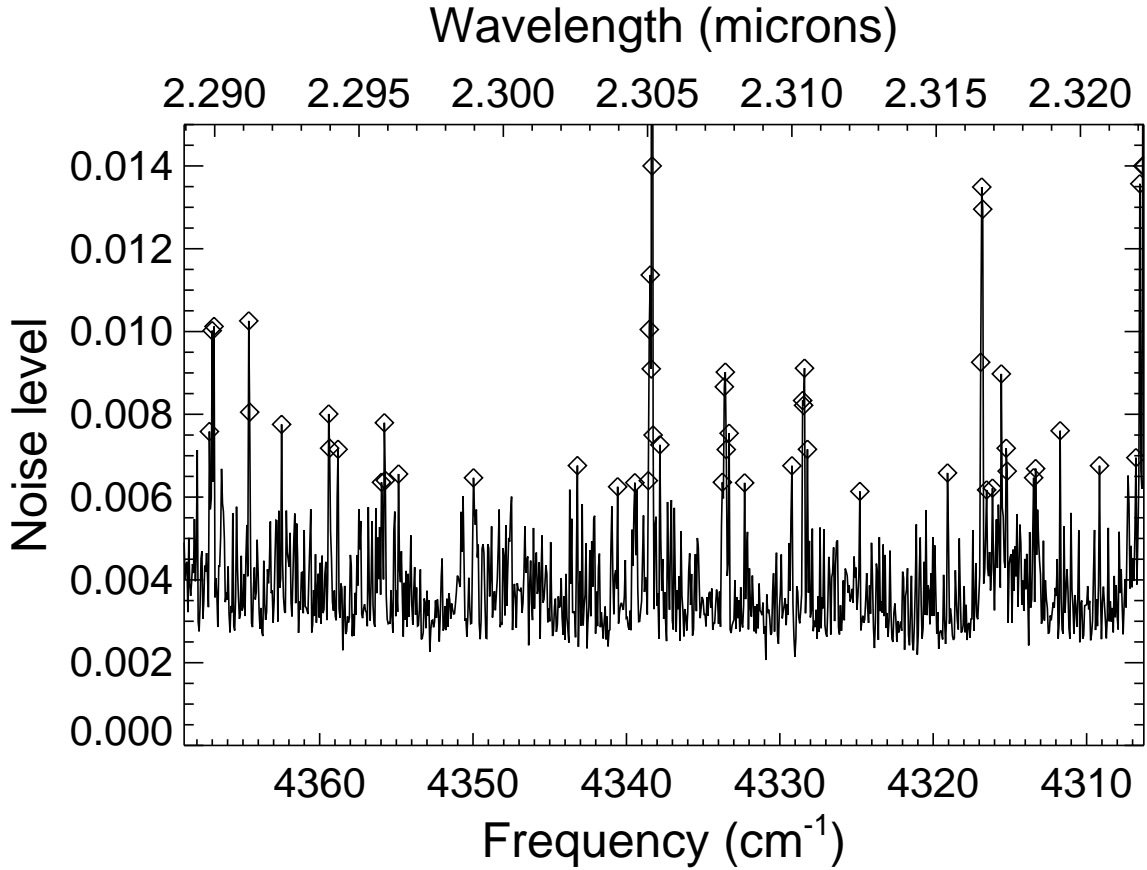}
\caption {Noise level $\sigma_{\lambda}$ achieved in echelle order 33
on 26 August 2002 {\em vs.} wavelength.  We compute $\sigma_{\lambda}$
from the variation of residual spectral intensity, $r_{\lambda t}$,
with $t$ (time) varying.  Diamonds indicate wavelengths not utilized
in subsequent analysis, due to the increased noise level near telluric
lines (compare to Figure 2). \label{fig3}}
\end{figure}
\clearpage

\begin{figure}
\plotone{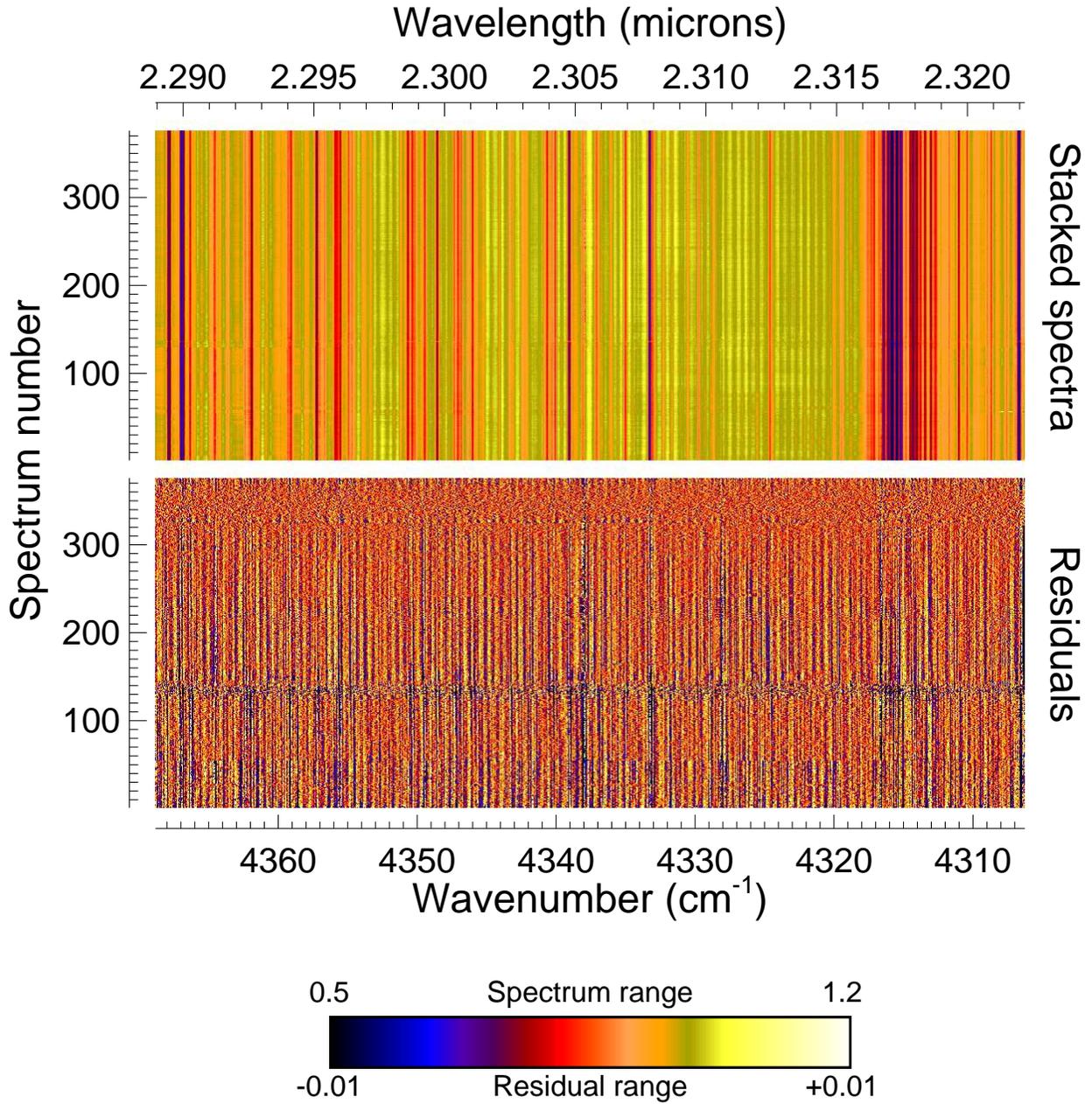}
\caption{Top: Wavelength-calibrated and airmass corrected spectra
for echelle order 33 on 26 August 2002.  Bottom: Spectral residuals
($r_{\lambda t}$) after digital and Fourier filtering. The color bar
shows the range of displayed intensities for the spectra and
residuals. \label{fig4}}
\end{figure}
\clearpage

\begin{figure}
\epsscale{0.6}
\plotone{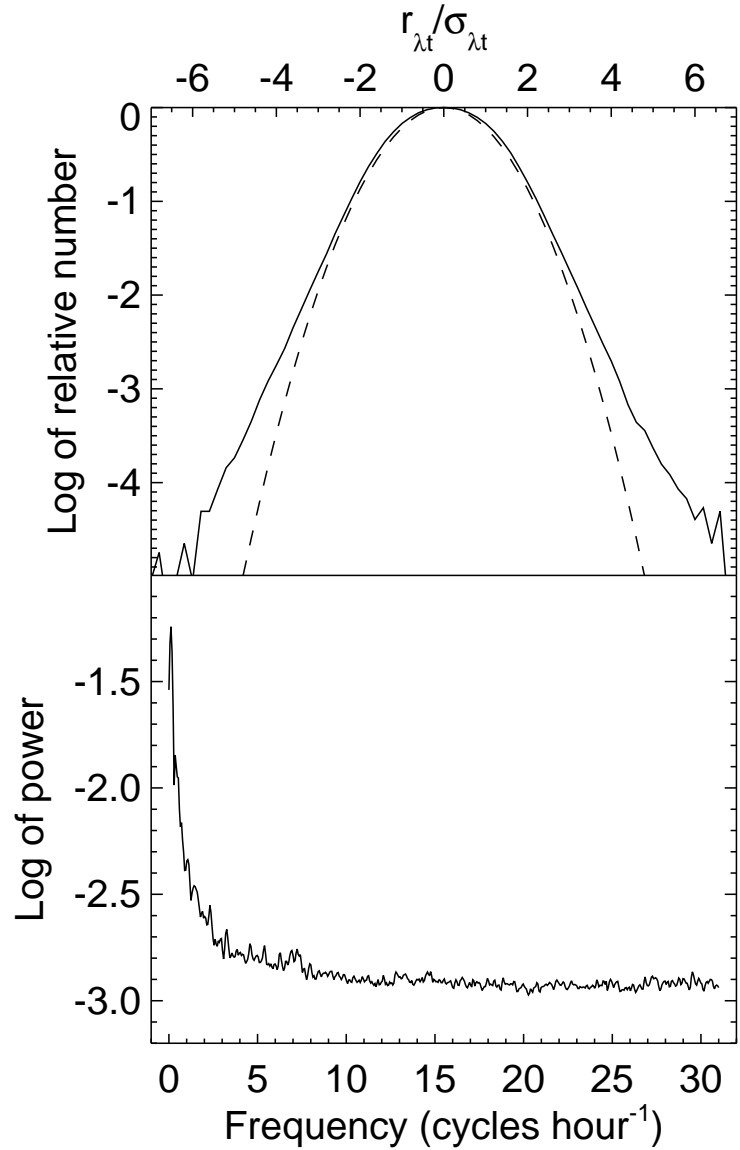}
\caption{Top: distribution of residual values (solid line) compared to a
Gaussian normal error distribution (dashed line). The bin size is 0.2
in ${r_{\lambda t}}/{\sigma_{\lambda t}}$. Note log scale used for the
ordinate. Bottom: average power spectrum of the residuals,
showing the increase in noise at low temporal frequencies.  This
average power spectrum has been normalized to a total power (area
under the curve) = unity.\label{fig5}}
\end{figure}

\begin{figure}
\epsscale{0.7}
\plotone{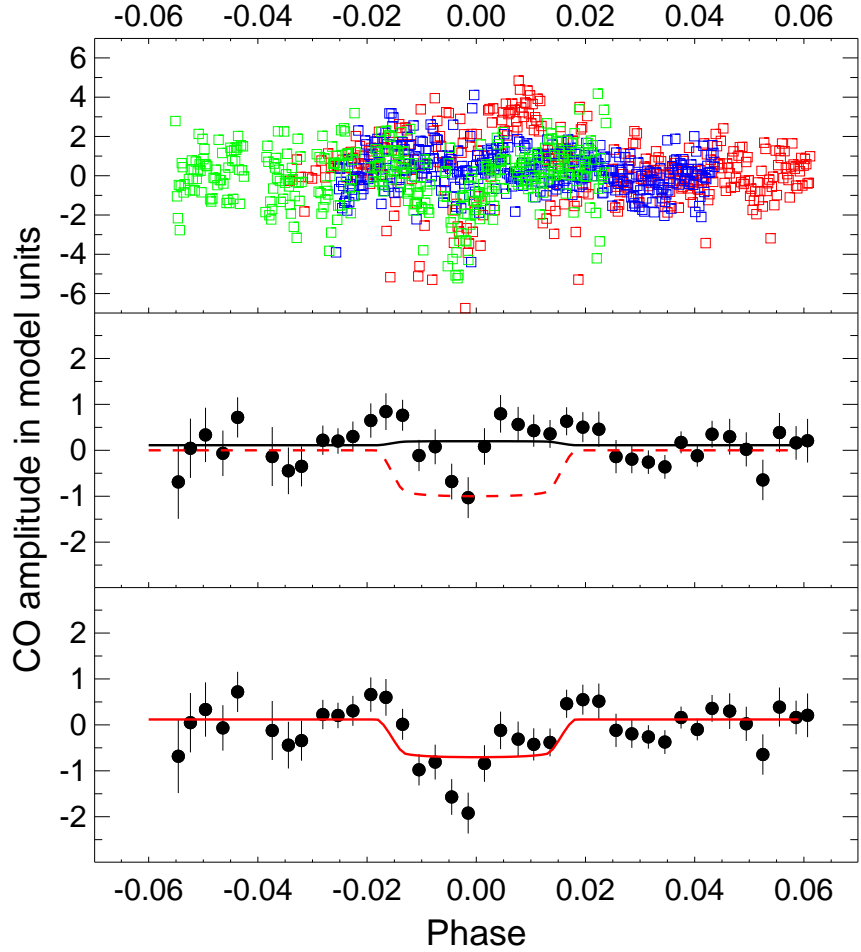}
\caption{Results for the carbon monoxide transit depth for the
fiducial model, having cloud tops at 37 mbar.  Top: Retrieved CO
amplitudes for all 1077 spectra over 3 observed transits (error bars
supressed), red = 19 August, blue = 26 August, green = 2 September.
Middle: Data binned by orbital phase (bin width 0.003), with error
bars.  The black solid line is the least squares fit of the transit
curve, having depth $-0.09 \pm 0.14$ in units of the fiducial model
(i.e., no transit CO signal). The red dashed line is a forced fit of
transit depth of unity, i.e. CO absorption during transit equal to the
fiducial model.  Bottom: The recovery of a fake signal input with CO
transit depth of unity, and recovered at depth $0.82 \pm 0.14$.
Phase=0 is the center of transit. \label{fig6}}
\end{figure}

\begin{figure}
\epsscale{0.8}
\plotone{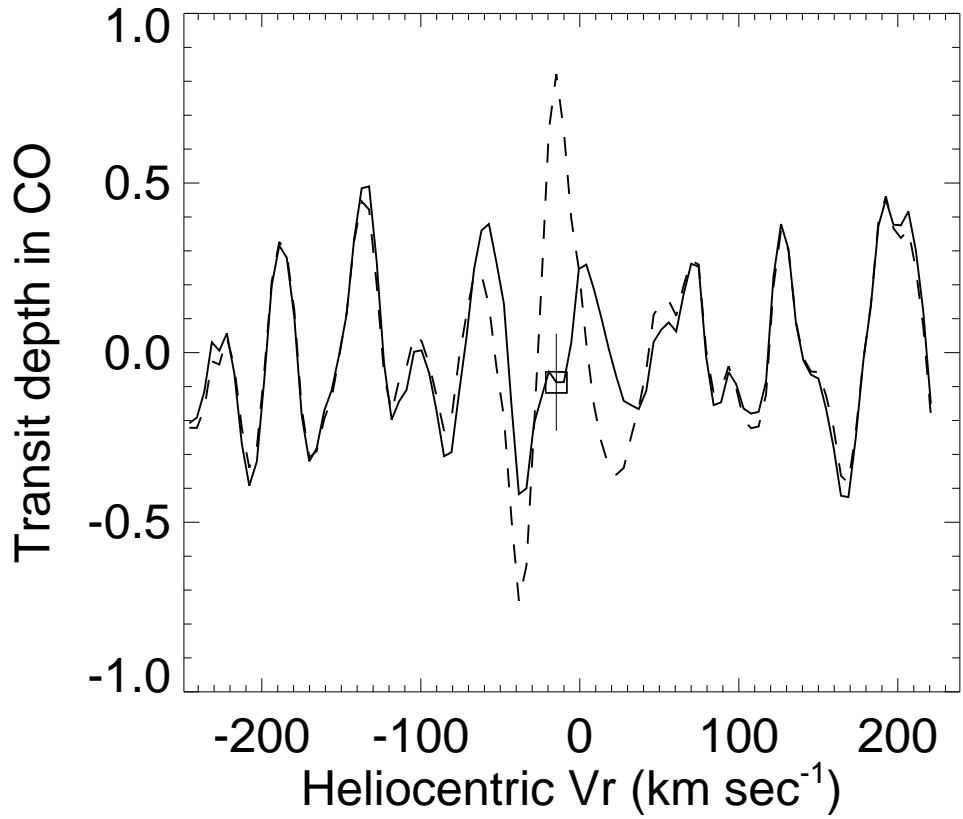}
\caption{Transit depth from our analysis as a function of the
heliocentric radial velocity ($V_r$) of the HD\,209458 center of mass.
The result from the fiducial model is plotted as a single point, with
error bar from the fit.  The dashed line shows recovery of the fake
signal at the correct $V_r = -14.8$ km sec$^{-1}$. \label{fig7}}
\end{figure}
\clearpage

\begin{figure}
\plotone{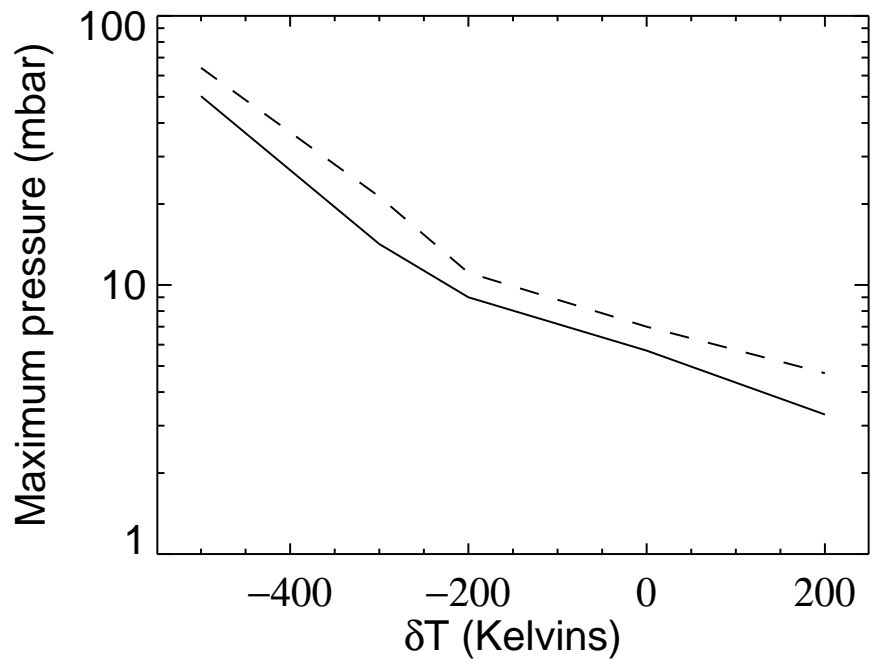}
\caption{Maximum cloud top pressure required to obtain consistency
with our observations as a function of the temperature perturbation 
($\delta T$) added to the \citet{sud03} model.  The dashed line gives 
the result of setting the CO mixing ratio to zero on the planet's 
anti-stellar hemisphere. \label{fig8}}
\end{figure}

\clearpage

\end{document}